\documentclass[
reprint,
superscriptaddress,
 amsmath,
 amssymb,
 aps,
 prd,
 floatfix,
 fleqn,
]{revtex4-2}
\usepackage[colorlinks=true,citecolor=blue,filecolor=blue,linkcolor=blue,urlcolor=blue]{hyperref}
\usepackage[dvipsnames]{xcolor}
\usepackage{graphicx}
\usepackage{multirow}
\usepackage{lineno}
\usepackage[title]{appendix}

\begin{document}

\title{Detector signal characterization with a Bayesian network in XENONnT}
\newcommand{\bologna}{\affiliation{Department of Physics and Astronomy, University of Bologna and INFN-Bologna, 40126 Bologna, Italy}}
\newcommand{\chicago}{\affiliation{Department of Physics and Kavli Institute for Cosmological Physics, University of Chicago, Chicago, IL 60637, USA}}
\newcommand{\coimbra}{\affiliation{LIBPhys, Department of Physics, University of Coimbra, 3004-516 Coimbra, Portugal}}
\newcommand{\columbia}{\affiliation{Physics Department, Columbia University, New York, NY 10027, USA}}
\newcommand{\lngs}{\affiliation{INFN-Laboratori Nazionali del Gran Sasso and Gran Sasso Science Institute, 67100 L'Aquila, Italy}}
\newcommand{\mainz}{\affiliation{Institut f\"ur Physik \& Exzellenzcluster PRISMA$^{+}$, Johannes Gutenberg-Universit\"at Mainz, 55099 Mainz, Germany}}
\newcommand{\heidelberg}{\affiliation{Max-Planck-Institut f\"ur Kernphysik, 69117 Heidelberg, Germany}}
\newcommand{\munster}{\affiliation{Institut f\"ur Kernphysik, Westf\"alische Wilhelms-Universit\"at M\"unster, 48149 M\"unster, Germany}}
\newcommand{\nikhef}{\affiliation{Nikhef and the University of Amsterdam, Science Park, 1098XG Amsterdam, Netherlands}}
\newcommand{\nyuad}{\affiliation{New York University Abu Dhabi - Center for Astro, Particle and Planetary Physics, Abu Dhabi, United Arab Emirates}}
\newcommand{\purdue}{\affiliation{Department of Physics and Astronomy, Purdue University, West Lafayette, IN 47907, USA}}
\newcommand{\rice}{\affiliation{Department of Physics and Astronomy, Rice University, Houston, TX 77005, USA}}
\newcommand{\stockholm}{\affiliation{Oskar Klein Centre, Department of Physics, Stockholm University, AlbaNova, Stockholm SE-10691, Sweden}}
\newcommand{\subatech}{\affiliation{SUBATECH, IMT Atlantique, CNRS/IN2P3, Universit\'e de Nantes, Nantes 44307, France}}
\newcommand{\torino}{\affiliation{INAF-Astrophysical Observatory of Torino, Department of Physics, University  of  Torino and  INFN-Torino,  10125  Torino,  Italy}}
\newcommand{\ucsd}{\affiliation{Department of Physics, University of California San Diego, La Jolla, CA 92093, USA}}
\newcommand{\wis}{\affiliation{Department of Particle Physics and Astrophysics, Weizmann Institute of Science, Rehovot 7610001, Israel}}
\newcommand{\zurich}{\affiliation{Physik-Institut, University of Z\"urich, 8057  Z\"urich, Switzerland}}
\newcommand{\paris}{\affiliation{LPNHE, Sorbonne Universit\'{e}, CNRS/IN2P3, 75005 Paris, France}}
\newcommand{\freiburg}{\affiliation{Physikalisches Institut, Universit\"at Freiburg, 79104 Freiburg, Germany}}
\newcommand{\napels}{\affiliation{Department of Physics ``Ettore Pancini'', University of Napoli and INFN-Napoli, 80126 Napoli, Italy}}
\newcommand{\nagoya}{\affiliation{Kobayashi-Maskawa Institute for the Origin of Particles and the Universe, and Institute for Space-Earth Environmental Research, Nagoya University, Furo-cho, Chikusa-ku, Nagoya, Aichi 464-8602, Japan}}
\newcommand{\laquila}{\affiliation{Department of Physics and Chemistry, University of L'Aquila, 67100 L'Aquila, Italy}}
\newcommand{\tokyo}{\affiliation{Kamioka Observatory, Institute for Cosmic Ray Research, and Kavli Institute for the Physics and Mathematics of the Universe (WPI), University of Tokyo, Higashi-Mozumi, Kamioka, Hida, Gifu 506-1205, Japan}}
\newcommand{\kobe}{\affiliation{Department of Physics, Kobe University, Kobe, Hyogo 657-8501, Japan}}
\newcommand{\ucla}{\affiliation{Physics and Astronomy Department, University of California, Los Angeles, CA 90095, USA}}
\newcommand{\kit}{\affiliation{Institute for Astroparticle Physics, Karlsruhe Institute of Technology, 76021 Karlsruhe, Germany}}
\newcommand{\tsinghua}{\affiliation{Department of Physics and Center for High Energy Physics, Tsinghua University, Beijing 100084, China}}
\newcommand{\ferrara}{\affiliation{INFN - Ferrara and Dip. di Fisica e Scienze della Terra, Universit\`a di Ferrara, 44122 Ferrara, Italy}}
\newcommand{\alsoatcoimbrapoli}{\affiliation{Coimbra Polytechnic - ISEC, 3030-199 Coimbra, Portugal}}
\newcommand{\alsoatuniheidelberg}{\affiliation{Physikalisches Institut, Universit\"at Heidelberg, Heidelberg, Germany}}
\newcommand{\alsoatroma}{\affiliation{INFN - Roma Tre, 00146 Roma, Italy}}

\author{E.~Aprile}\columbia
\author{K.~Abe}\tokyo
\author{S.~Ahmed Maouloud}\paris
\author{L.~Althueser}\munster
\author{B.~Andrieu}\paris
\author{E.~Angelino}\torino
\author{J.~R.~Angevaare}\nikhef
\author{V.~C.~Antochi}\stockholm
\author{D.~Ant\'on Martin}\chicago
\author{F.~Arneodo}\nyuad
% ...
\collaboration{XENON Collaboration}\email[]{xenon@lngs.infn.it}\noaffiliation

\noaffiliation

\date{\today}
\begin{abstract}
\newpage
\noindent We developed a detector signal characterization model based on a Bayesian network trained on the waveform attributes generated by a dual-phase xenon time projection chamber. 
By performing inference on the model, we produced a quantitative metric of signal characterization and demonstrate that this metric can be used to determine whether a detector signal is sourced from a scintillation or an ionization process. 
We describe the method and its performance on electronic-recoil (ER) data taken during the first science run of the XENONnT dark matter experiment. 
We demonstrate the first use of a Bayesian network in a waveform-based analysis of detector signals. 
This method resulted in a 3\% increase in ER \mbox{event-selection} efficiency with a simultaneously effective rejection of events outside of the region of interest. 
The findings of this analysis are consistent with the previous analysis from XENONnT, namely a background-only fit of the ER data.

\end{abstract}
\maketitle

\section{Introduction}

XENONnT is a dark matter \mbox{direct-detection} experiment currently operating at INFN Laboratori Nazionali del Gran Sasso in Italy. 
The experiment has a wide range of (astro)particle physics capabilities, including the search for weakly interacting massive particles (WIMPs), solar axions, and coherent \mbox{electron-neutrino} nucleus scattering from \textsuperscript{8}B solar neutrinos~\cite{Aprile_2020}.
XENONnT's physics program aims to make substantial progress on sensitivities through improvements in hardware, software, and analysis methods.
The XENON Collaboration recently reported a search for WIMPs~\cite{xenoncollaboration2023dark} and an analysis of \mbox{electronic-recoil} (ER) data~\cite{Aprile:2022vux} using the data collected during Science Run 0 (SR0).  
These new results relied on XENONnT's unprecedentedly low background radioactivity rates and its keV-scale energy threshold. 
To achieve the low intrinsic background rates in ultrarare event search experiments, large numbers of events not originating from the desired interactions in the target must be rejected; to this end, robust classification and characterization of detector signals are crucial for dark matter searches and other background-dominant processes, including solar axions, neutrino studies, and rare nuclear decays. 

During the past decade, the incorporation of machine learning techniques, particularly deep learning, has led to innovation within the field of (astro)particle physics~\cite{Radovic2018dip,Psihas_2020,Karagiorgi2022}. 
Advances in computer vision have made convolutional neural networks a common approach to \mbox{deep-learning} applications in (astro)particle physics for classification problems; see for instance~\cite{deOliveira:2015xxd,Komiske:2016rsd,Macaluso_2018,Abbasi:2021,Abbasi:2021,MicroBooNE:2020hho,Aurisano:2016jvx,DUNE:2020gpm,Abbasi:2021}.
Given the success of deep learning methods, applications of machine learning in direct dark matter searches with time projection chambers have seen increased attention~\cite{PhysRevD.106.072009}.
Recently, the use of Bayesian networks has been presented for event localization~\cite{Peters_2022} and for inference in the search for dark matter~\cite{50collaboration2023search}.

Motivated by such applications, we developed a Bayesian network for a waveform-based analysis of detector signals, where the waveform is defined as the shape of the signal observed by the photosensors as a function of time.
With this method, we aimed to quantify how alike a detector signal is to the models of scintillation (S1 signals) and ionization (S2 signals). 
S1 and S2 signals which can be explained by these models are defined as ``canonical'' in shape. 
We constructed a Bayesian network to quantify the extent to which a detector signal shape is canonical.
We evaluated how the model performs in classifying S1 and S2 signals, respectively, and compared the model performance to that of the baseline method of signal classification described in~\cite{Aprile:2022vux}. 
Then, we further applied this quality metric beyond signal classification as the primary detector signal quality feature in event selection. 
We present the first application of a \mbox{waveform-based} analysis of detector signals using a Bayesian network by analyzing the electronic-recoil data reported in~\cite{Aprile:2022vux} and obtain results that are in agreement with the original work.

The paper is organized as follows. 
In Sec.~\ref{sec:xenonnT} we provide an overview of the XENONnT experiment.
In Sec.~\ref{sec:signal_classification} we describe the use of a Bayesian network to study S1 and S2 signal classification. 
In Sec.~\ref{sec:signal_characterization} we apply the quantitative scores from the Bayesian network for selecting ER events, and we further present the results of this application. 
And finally in Sec.~\ref{sec:conclusions} we summarize the results and outline future work. 

\section{The XENON\lowercase{n}T Experiment}\label{sec:xenonnT}

The XENONnT detector is a dual-phase xenon time projection chamber (TPC) with an active target mass of 5.9~tonnes of liquid Xe (LXe). 
Detailed information regarding XENONnT detector conditions, systems, and subsystems can be found in \cite{Aprile_2017,Aprile2017_1t_kr,Aprile_2014_muon,Aprile_2022_rd,Murra_2205,Antochi_2021,Aprile:2022vux}.

The working principle of the detector can be described as follows: when a particle interacts within the LXe, the energy transferred to the target excites and ionizes the atoms. 
The excitation of Xe atoms creates dimer states, which then decay promptly by emitting vacuum ultraviolet photons. 
These photons are detected with photomultiplier tube (PMT) arrays at the top and bottom of the TPC. 
The prompt scintillation signal of an interaction is called an S1 signal. 
Simultaneously, the ionization of Xe atoms at the interaction site liberates atomic electrons. 
Some of these electrons recombine with nearby ions to create excited states, which then contribute to the S1 signal.
The remaining electrons are drifted upward to a liquid-gas interface by an applied electric field between a cathode at the bottom of the detector and a gate electrode at the top of the liquid. 
A thin gaseous Xe layer above the liquid acts as an amplification region, where an anode accelerates the electrons into the gas, generating a proportional electroluminescence signal.
This secondary ionization-induced signal is called an S2 signal, and its size is proportional to the number of extracted electrons. 

For any detector signal, the electronic pulses from different PMTs are clustered in time, forming a waveform, as discussed in Sec.~\ref{sec:data}. 
Given the stochastic nature of the prompt and secondary scintillation processes, each waveform will have a unique shape and size, with a typical S1 signal having a much narrower time profile than a typical S2 signal, owing to the nature of their production mechanisms. 
An event is formed from a pair of S1 and S2 detector signals from a single-scatter interaction. 
The interaction 3D position and energy deposited as ionization and scintillation light is reconstructed for each event. 

The XENONnT SR0 electric field configuration presented new low-energy analysis opportunities due to a higher light yield and thus a lower energy threshold, but also introduced new challenges for event reconstruction and background mitigation, including lower electron production and longer electron drift time.
For more information on the electric field configuration in XENONnT SR0, see~\cite{Aprile:2022vux}. 

Challenges to a sensitive rare-event analysis in this case include effectively rejecting interactions that occur in the top gas region of the TPC, mostly electronic recoils from gamma rays produced by radioactive contamination in the detector material.
Gaseous xenon (GXe) interactions  can be mistakenly reconstructed to have occurred within the LXe region of the TPC, requiring effective gas-event mitigation for analysis. 
Ionization signals from GXe interactions have characteristic shapes and can thus be identified by analyzing the ionization signal waveform.

Additionally, the rate of accidental coincidences (ACs), which occur when two detector signals classified as S1 and S2 signals do not originate from the same single-scatter interaction within the TPC, increases with a longer electron drift time and errors in detector signal classification.
These challenges require careful approaches both to classification of S1 and S2 signals and to event selection based on the characteristics of detector signals comprising an event, to reduce events outside of the region of interest (ROI) for a given analysis.

\section{Signal Classification} \label{sec:signal_classification}

Current and previous S1 and S2 signal classification algorithms employed in direct dark matter detection experiments with LXe TPCs rely on manually analyzing and optimizing a decision boundary using the size (area) and the width of the signal~\cite{XENON:2019ykp,LUX:2017bef,PandaX-4T:2021bab,LUX-ZEPLIN:2022qhg}. 
In XENONnT SR0, a waveform was classified as an S1 (S2) signal if its waveform rises sufficiently quickly (slowly) and has at least 3 (4) contributing PMTs~\cite{Aprile:2022vux}. 
This classification method lacks the ability to provide an informative classification confidence metric, or score, owing to its being deterministic.
Therefore, further detector signal quality selections must be applied to reduce detector signal misclassification rates.
To overcome the limitations of using this detector signal classification method, we developed a waveform-based classification model based on a Bayesian network, which produces continuous outputs that can be used in an informative way.

\subsection{Input data} \label{sec:data}
The training and evaluation data for this work consist of a combination of simulated and measured data, primarily simulated ER interactions ranging from [0.75,~200]~keV. 
The simulated data were generated using XENONnT’s waveform simulator package,  \verb|WFSim|~\cite{peter_gaemers_2022_6783261}, and event-building software, \verb|straxen|~\cite{joran_r_angevaare_2022_6821967}. 
Light and charge yields used in \verb|WFSim| are computed using \verb|NEST|~\cite{nest_v221, nestpy_v149}. 

In \verb|WFSim|, the light yield for an interaction with a given energy is used to generate S1 signals before computing the photons' arrival times at the PMTs. 
The light yield accounts for scintillation of the Xe atoms, \mbox{electron-ion} recombination, and the singlet-to-triplet fraction of excited states. 
The observed \mbox{light-yield} probability distribution is computed for each PMT based on an S1 \mbox{light-collection} efficiency (LCE) map. 
The simulation process for S2 signals differs because electrons are drifted toward the liquid-gas interface, during which electronegative impurities may reduce the number of electrons arriving at the liquid-gas interface.
The arrival times of the electrons at the liquid-gas interface are computed based on electron drift and diffusion models, followed by calculating the photon timing of the electroluminescence of individual electrons. 
Then, based on the S2 LCE map, the \mbox{charge-yield} probability distribution is computed for each PMT. 

Subsequently, for both S1 and S2 signal simulations, once the arrival times of photons are computed, the currents in the PMTs are simulated using a model of the PMT and digitizer response with sampling-time resolution of 10~ns~\cite{XENON:2022vye}. 
Then, using \verb|straxen|, PMT signals are scanned for PMT ``hits" above a threshold.
PMT hits are grouped iteratively into clusters with adjacent hits within a \mbox{700-ns} time window, forming a waveform that can subsequently be classified as an S1, S2, or ``unknown" by \verb|straxen|. 
Hits with no neighbors in the time window are treated as isolated hits.
In XENONnT simulations and SR0 data taking, every waveform is saved with a fixed number of 200 samples. 
Sample duration in XENONnT can vary from a few tens of nanoseconds for an S1 signal and up to few hundreds of nanoseconds for an S2 signal, to accommodate the full waveform within the fixed 200-sample interval. 

We augmented the training data in the low-area range where signal classification is particularly challenging in SR0.
To this end, we simulated a set of S2-only signals with [1,~10]~electrons (at very low ER interaction energies, S1 signals are rarely observable and are below the energy threshold of 1~keV used in the low-energy ER analysis). 
Additionally, a very pure ($>$99.9\%) measured sample of single-electron S2 signals, produced from a highly emitting electron source near the gas region, was added to ensure that experimental effects at these low areas were represented in the training sample despite any limitations in modeling at these low areas. 
The measured \mbox{single-electron} data represented 2.5\% of the training data, the cause of which was a short between the bottom screening and cathode electrodes which produced intermittent, localized bursts of single electrons that could be tagged via their position to the location of their source.
In total, the simulation and measured data used in training and evaluation consisted of 10$^6$ waveforms with ground-truth labels. 

The training dataset was composed of true S1 or S2 waveforms from single-scatter interactions within the detector's \mbox{4-tonne} fiducial volume, which is defined in~\cite{Aprile:2022vux}.
In this work, we define \mbox{out-of-distribution} detector signals as any types of waveforms which were not included in the training set for the model.
This includes S1 or S2 signals with noncanonical shapes and interactions in gas, where both can be vetoed during analysis based on their characteristics, like waveform shape, as will be demonstrated later.

\begin{figure}
    \centering
    \vspace{0mm}
    \includegraphics[width=\linewidth]{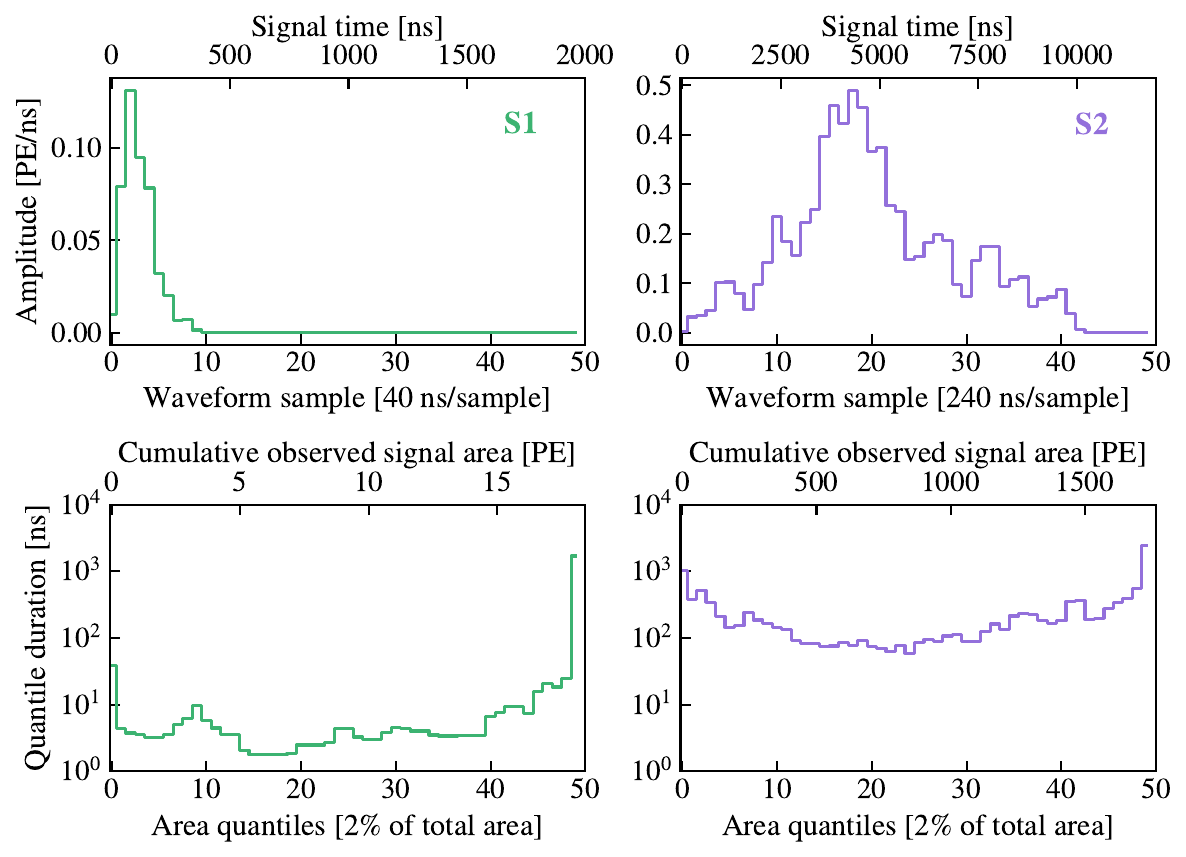}
    \caption{Individual examples of an S1 signal (green) and S2 signal (purple) from simulated data used as input for training.
    The S1 signal example has a total area of 18~PE and duration of 2000~ns; the S2 signal area is 1766~PE and duration is 12000~ns.
    Top:  waveforms, downsampled to 50~samples total, with the elapsed time of the signal in the secondary \textit{x} axis, illustrating the different temporal profiles between a typical S1 and S2 signal. 
    Bottom:  quantiles, 50 total, with relative total observed area in the secondary \textit{x} axis.
    }
    \label{fig:waveform_quantile}
\end{figure}

The top panels of Fig.~\ref{fig:waveform_quantile} show simulated S1 and S2 waveforms downsampled to 50~samples, with an absolute amplitude in units of [PE/ns].
The waveform samples represent the light collected in a certain window of time. 
The bottom panels of Fig.~\ref{fig:waveform_quantile} show the second component of the input to the algorithm, quantiles with 50~elements. 
The quantiles represent the amount of time elapsed for a given fraction of the total waveform area to be observed --- for instance, if considering 50 quantiles, each quantile is the duration of time elapsed for an additional 2\% of the total waveform area to be observed. 
The waveforms and quantiles are transformations of each other, where the quantiles explicitly contain time-profile information, and the waveforms explicitly contain signal-size information. 
These discretized samples of the data are treated as individual attributes, as explained in Section~\ref{sec:sec_NBC}. 

\subsection{Classification with a naive Bayes classifier} \label{sec:sec_NBC}

A naive Bayes classifier (NBC) is a type of Bayesian network that uses a simple graph-based representation to compactly encode a complex, \mbox{high-dimensional} distribution~\cite{VermaPearl1988, Pearl1988}.
For a comprehensive explanation of Bayesian networks, see e.g.~\cite{Koller2009}.
 
\begin{figure}
    \vspace{0mm}
    \centering
    \includegraphics[width=\linewidth]{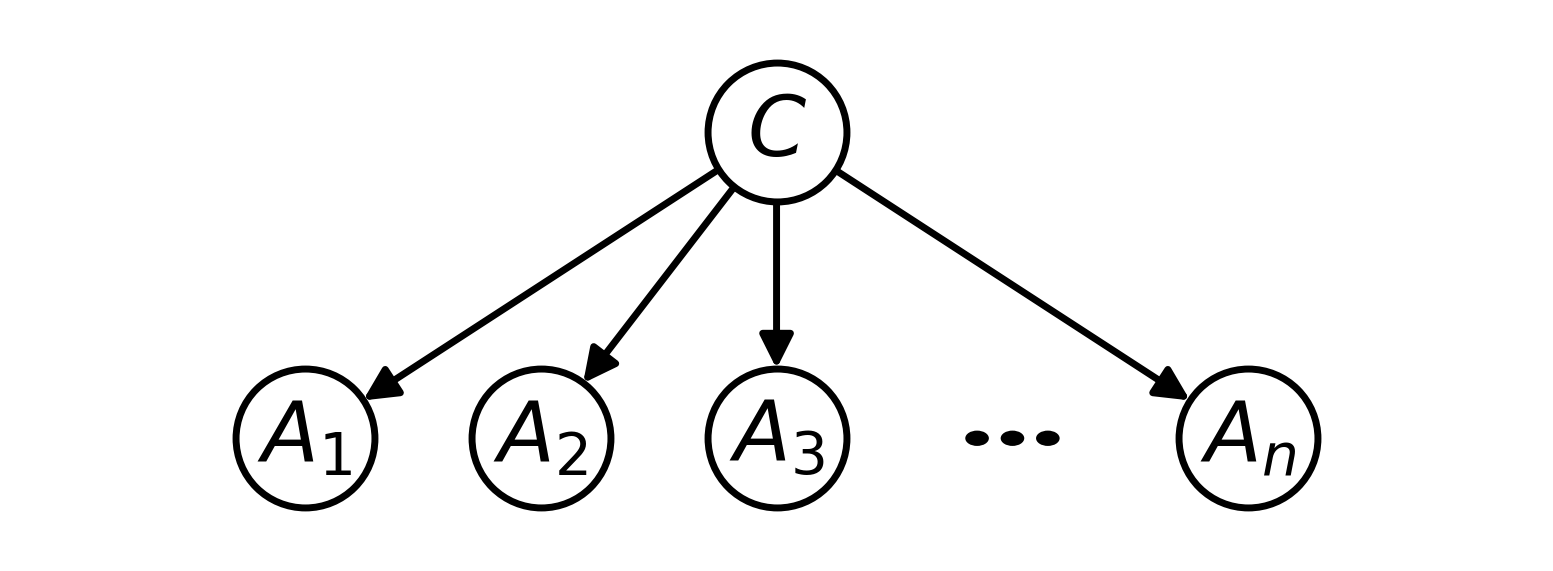}
    \caption{Graph structure of a naive Bayes classifier. 
    Class node ($C$, either S1 or S2) is assumed to directly influence the value of each attribute node ($A_i$). 
    Each attribute node is conditionally independent from all other attribute nodes, given the class node.}
    \label{fig:naive_bayes}
\end{figure}

The graph structure of an NBC is shown in Fig.~\ref{fig:naive_bayes}, where each circle in the graph is called a node, which represents a variable, discrete or continuous, that depends on a stochastic process. 
This is known as a random variable. 
The arrows between nodes denote dependence between random variables, and the direction denotes causality.
To avoid a variable depending upon itself, there can be no cyclic paths in the graph.
The class node, $C$, is the parent node of the set of attribute nodes, $\vec{A} = \{ A_1,\ A_2,\ \dots,\ A_{n} \}$. 

The graph structure of an NBC implies that all attributes are directly \textit{dependent} on the class, and conditionally \textit{independent} from each other, given the class. 
In this work, the attributes denote quantiles and waveform elements of time-series data, as depicted in Fig.~\ref{fig:waveform_quantile}.
The independence assumption is naive in this case because the attributes, being time-series data, are not truly independent.
The class node maps each attribute to the signal classification, either S1 or S2. 
In practice, the set of possible values that the random variable $C$ can take is $\{ 0,\ 1 \}$, which are indices corresponding to the discrete classes.

The quantiles and waveform elements are continuous values. 
Therefore, the range of possible values each attribute can take on must be either parameterized or discretized.
For this work, we discretize the attribute's values by binning and assigning indices corresponding to the bin.
The Bayesian blocks method performs discretized, nonuniform binning of a continuous variable based on the observed values for that random variable attribute, as described in Sec.~3.1 of~\cite{Scargle_2013}.
The set of possible values that an attribute can take is $\{ 0,\ \dots,\ m \}$, where $m$ is the number of bins defined by the Bayesian blocks method. 

It follows that each entry in the full joint distribution over all of the random variables in this graph structure is defined as:
\begin{equation}\label{eq:bayes_joint}
    P(C,\ A_{1},\ \dots,\ A_{n}) \ \propto \ P(C) \prod_{i=1}^{n} P(A_i\ |\ C ),
\end{equation}
where $P(C)$ is the prior distribution over the signal classes and $P(A_i\ |\ C )$ is the local joint probability distribution of the $i$th attribute conditioned on the signal class.
For this work, we chose to use a flat prior on the class distribution, $P(C)$, meaning a 50/50 split of S1 and S2 signals.
This is a logical choice of prior for the ER physics analysis presented here, which requires both an S1 signal and an S2 signal for an event to be considered. 
Regardless, the choice of prior was not found to significantly affect the performance of the classifier.

The probability distribution of the attributes conditioned upon the signal class is learned from the training data for each attribute. 
Once the conditional probability distributions are learned, the probability of a waveform belonging to each class can be inferred using Bayes' rule.
This is known as a probability query and is defined as:
\begin{multline} \label{eq:bayes_rule}
    P(C\ |\ A_{1} = a_1,\ \dots,\ A_{n} = a_n) = \\ \frac{P(C) \prod\limits_{i=1}^{n} P(A_i = a_i\ |\ C )}{P(A_{1} = a_1,\ \dots,\ A_{n} = a_n)}\ ,
\end{multline}
where the result of the query is the posterior probability distribution over the values of $C$, conditioned on the observed values of the attribute nodes, $\vec{a} = \{ a_1,\ a_2,\ \dots,\ a_{n} \}$.
Note that the denominator does not depend on the value of $C$.
In practice, the natural logarithm of the posterior probability is calculated to avoid computational loss of precision.

The number of attributes used in evaluation and subsequently in Sec.~\ref{sec:signal_characterization} was chosen by training for S1/S2 signal classification using multiple attribute options and selecting the choice of attributes with optimal classification performance.
In this work, we found that 100 attributes comprising 50~waveform~samples and 50~quantiles, as shown in Fig.~\ref{fig:waveform_quantile}, had the highest classification performance of those studied. 

We found the benefits of using an NBC for signal classification to be threefold. 
First, the Bayesian classification approach is intuitive and interpretable.
Second, the NBC structure has been shown to be effective for classification even in cases where there are strong dependencies between the attribute nodes \cite{NBC:article}, with the advantages of being faster to learn, faster to query, and smaller to store in memory than a graphical model that includes complex dependencies among attributes.
Finally, the output of a probability query of the network is informative about the network's confidence in a signal belonging to each class; this capability is not present in current and previous S1 and S2 signal classification algorithms~\cite{XENON:2019ykp,LUX:2017bef,PandaX-4T:2021bab,LUX-ZEPLIN:2022qhg}. 

\subsection{Naive Bayes classifier performance}
The NBC was constructed and trained by building upon the scientific \verb|Python| software stack~\cite{python,Harris2020,SciPy2020}.
The trained model was evaluated using 50\% of the labeled data, which were not used in training.   
For each signal, the most probable class, S1 or S2, can be decided by performing a probability query for each class and choosing the most probable class. 

The NBC can be evaluated as a deterministic classifier by taking a static decision boundary on the posterior distribution, in this case $P(C=S1|\vec{A})=P(C=S2|\vec{A})=0.5$. 
The classification performance of the NBC is shown in Table~\ref{table:classification}. 
The NBC, even given the independence assumptions described in Sec.~\ref{sec:sec_NBC}, outperformed the \verb|straxen| classification, which correctly classified 99.974\% of S1 signals and 99.907\% of S2 signals.
We ascribe this improvement to the NBC having sufficient parameters to describe the conditional probability distributions encoded in the attributes.

\begin{table}
\begin{center}
\begin{tabular}{ |c|c|c|}

\multicolumn{3}{c}{Naive Bayes Classifier} \\
\hline
& True S1 & True S2\\
\hline
Predicted S1 & 99.999~$\pm~0.001~\%$& 0.003~$\pm~0.001~\%$ \\ 
Predicted S2 & 0.001~$\pm~0.001~\%$& 99.997~$\pm~0.001~\%$ \\ 
\hline
\end{tabular}
\caption{Performance results on evaluation data of the naive Bayes classifier described in Sec.~\ref{sec:sec_NBC} for detector signal classification. 
True S1 (S2) denotes the labeled S1 (S2) signal populations in the validation dataset. 
}
\label{table:classification}
\end{center}
\end{table}

\subsection{Naive Bayes classifier score as a signal characterization metric}
In addition to the NBC being an effective deterministic classifier, the values output from the probability queries were found to be informative about the characteristics of the S1 or S2 signal's shape. 
Due to violation of the independence assumptions implied by the NBC structure, as well as the use of an uninformative prior, in this case the probability query returns a ``score" rather than a reliable probability.
Nevertheless, the scores from the NBC can be used to characterize each signal beyond the task of deterministic S1/S2 signal classification. 
We define the NBC score as the natural logarithm of the ratio of the calculated probability queries of the class, $C$, having the value S1 or S2, given the values of the attributes: 
\setlength{\mathindent}{0cm} 
\begin{multline} \label{eq:NBCscore}
    \text{NBC score} = \ln{(P(C=\text{S}1\ |\ A_{1} = a_1,\ \dots,\ A_{n}  = a_n)})\\ - \ln{(P(C=\text{S}2\ |\ A_{1} = a_1,\ \dots,\ A_{n} = a_n))} \ .
\end{multline} 
This NBC score quantifies the extent to which the model favors the signal being of either canonical scintillation origin (S1-like) or canonical ionization origin (S2-like).
A detector signal with a large positive value for NBC score has an S1-like shape, while a large negative value for NBC score would imply an S2-like shape. 
Detector signals with NBC scores near 0 are neither \mbox{S1-like} nor \mbox{S2-like}.
Each event, which consists of an S1 and S2 signal pair, will have an S1 NBC score, referring to the NBC score of the S1 signal, and an S2 NBC score, referring to the NBC score of the S2 signal.

To illustrate the utility of the NBC score beyond deterministic classification, we considered the calibration source $^{83\text{m}}$Kr, which decays via two ER-inducing internal conversion steps, first depositing 32.1~keV and then 9.4~keV, with a half-life of the latter decay of 154~ns~\cite{RUBY1971321}.
$^{83\text{m}}$Kr is used as a standard candle for calibration in XENONnT and other xenon TPCs due to its distinct properties, including this prompt two-step decay. 
In practice, the two decays are often merged into a single S1 signal due to the intermediate state's short half-life being on the same order of magnitude as an S1 signal's total width, with the much wider S2 signals almost always being merged into a single S2 signal. 
Therefore, correctly distinguishing the separated S1 signals from the merged S1 signals in $^{83\text{m}}$Kr data is critical for energy response studies, which rely on properly characterizing S1 signals as being merged or separated.

\begin{figure}
    \centering
   \includegraphics[width=\linewidth]{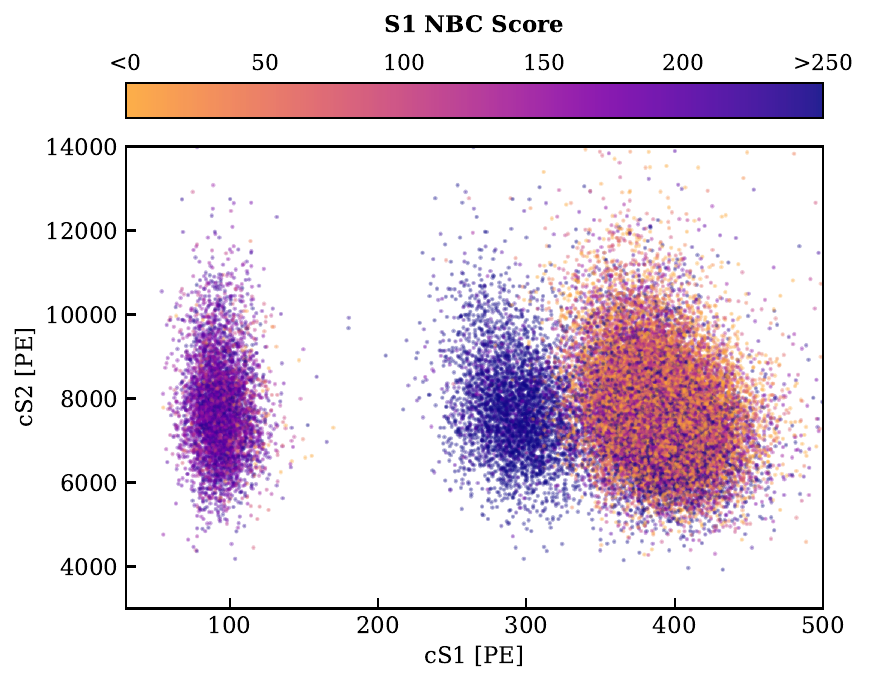}    
    \caption{$^{83\text{m}}$Kr calibration data corrected S2 area vs. corrected S1 area (cS2, cS1). 
    Corrections on area are calculated to normalize detector effects that vary across time and space, see~\cite{Aprile:2022vux} for details.
    The color scale indicates the S1 signal's NBC score (Eq.~\ref{eq:NBCscore}). 
    The rightmost population contains the merged (\mbox{41.5-keV}) S1 signals, and thus is shown to have an NBC score that is less canonically S1-like, owing to the true underlying physical process being a merging of two S1 signals.
    The \mbox{32.1-keV} (middle) and \mbox{9.4-keV} (left) populations have S1 NBC scores which are more canonically S1-like.
    The S2 signals for the \mbox{32.1-} and \mbox{9.4-keV} are merged into a single S2 signal; thus, all three populations of S1 signals shown have equivalent-sized S2 signals.
    }
    \label{fig:kr83m_nbc}
\end{figure}

Figure~\ref{fig:kr83m_nbc} demonstrates a case where NBC scores can be used to isolate detector signals which are noncanonical in shape, i.e., not truly isolated S1 signals produced within the fiducial volume of the detector.
Examples of different waveform topologies are shown as additional material in the Appendix.
Based on the results of this test case, the NBC scores were concluded to be a useful metric for reducing temporal overlap of signals. 
Furthermore, in the following section, we demonstrate that, due to its robustness against out-of-distribution samples, applying the NBC scores in event selection is suitable for physics analyses where one wishes to reject S1 and S2 signals of noncanonical shapes.

\section{Event selection using the NBC scores} \label{sec:signal_characterization}
In this section we describe the application of signal characterization in event selection based on NBC scores to the XENONnT SR0 ER dataset. 

Most recorded signals in a dark matter detector science run, as in XENONnT’s SR0 campaign, are not derived from single-scatter recoil interactions in the fiducial volume. 
They are primarily due to mislabeled detector signals, detector signals from interactions in the gas which are misplaced within the fiducial volume, multiple detector signals which are merged together in processing, and detector signals grouped from lone hits from multiple PMTs, such as dark counts.
These signals can contribute to events that are defined as being outside of the ER ROI.
Conversely, events within the ER ROI are defined as single-scatter ER events within the energy range of interest and occurring within the fiducial volume. 

Because the NBC was trained on S1 and S2 signals generated from ER interactions within the fiducial volume, events outside of the ER ROI, relative to ER events within the ROI, will have neither strongly S1-like or S2-like NBC scores for their S1 and S2 signals, respectively.
The NBC score is a single metric that represents the characteristic of the full waveform shape. 
Given this feature of the NBC scores, we evaluated their ability to be used in place of several low-dimensional quality selection parameters that were used in the SR0 analysis in XENONnT as \mbox{signal-quality} selection criteria~\cite{Aprile:2022vux}.

We calculated the efficiency of applying the NBC scores to select ER events and the effectiveness of removing events outside the ER ROI using calibration data from $^{220}$Rn and $^{37}$Ar injections in XENONnT.
The $^{220}$Rn decay chain produces $^{212}$Pb, a uniform $\beta$-emitting source of ER events across the \mbox{[1-140]-keV} energy region in this analysis. 
The $^{220}$Rn data were used to define signal selection efficiency across the ER energy spectrum~\cite{PhysRevD.95.072008}.
$^{37}$Ar decays by electron capture into $^{37}$Cl, producing an electron vacancy in either the \textit{K}, \textit{L}, or \textit{M} shell. 
In particular, the \textit{K}-shell vacancy (90.2\% branching ratio) being filled by an electron rearrangement deposits a total energy of 2.82~keV~\cite{Boulton_2017}, which was useful in this study for calibrating ER event acceptance near the threshold of detection with high statistics and minimal contamination. 

We calculated NBC scores for both the S1 signal and S2 signal of each event in the calibration data.
Each event's S1 (S2) signal selection depends upon a decision boundary between the NBC score of S1 (S2) signals and the S1 (S2) observed signal area.
The optimal decision boundaries used in this selection were determined using the $^{220}$Rn calibration dataset and chosen to isolate noncanonical detector signals.
It is noteworthy that the selection can be made more strict or loose depending on the analysis application.
We then evaluated both the efficiency of accepting ER-like events, and the effectiveness at rejecting events outside of the ER ROI, using the NBC score boundaries that were imposed on S1 and S2 signals contributing to events reconstructed within the fiducial volume. 

The selection of ER events in this Bayesian network-based method differs from the previous work in~\cite{Aprile:2022vux} by reducing several signal-quality selection criteria applied in~\cite{Aprile:2022vux} to only two: one selection on the S1 signal's NBC score, and one selection on the S2 signal's NBC score. 
Both methods further apply identical additional selection criteria to remove remaining multiple-scatter events, mispaired S1 and S2 signals, and accidental coincidence events. 

\begin{figure}
    \vspace{1mm}
    \centering
    \includegraphics[width=\linewidth]{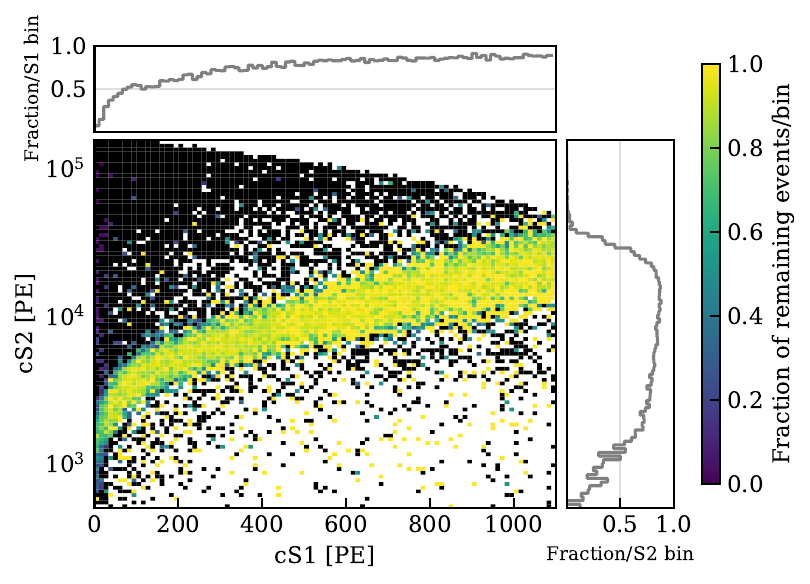}
    \caption{Performance of the S1 NBC and S2 NBC selection criteria on $^{220}$Rn calibration data, within the fiducial volume. 
    A bin colored black indicates the fraction of remaining events to be exactly 0.
   Adjacent panels show the fraction of remaining events per bin, projected along the cS1 and cS2 axes.
   Events outside of the ROI are effectively targeted by the S1 and S2 NBC selection criteria. 
   Above the ER band, gaseous S2 signals are misreconstructed into events within the fiducial volume, becoming the primary source of background events. 
   Following the S1 and S2 NBC selection criteria, selections to target multiple-scatter events, mispaired S1 and S2 signals, and accidental coincidences are applied, which remove the remaining events outside of the ER ROI. 
   }
    \label{fig:cs1_cs2_rn_bayes_eff}
\end{figure}

Figure~\ref{fig:cs1_cs2_rn_bayes_eff} shows the S1 and S2 NBC selection performance on $^{220}$Rn calibration data.
The bulk of the events removed from the $^{220}$Rn calibration data contain gaslike S2 signals and predominantly lie above the band of ER events in Fig.~\ref{fig:cs1_cs2_rn_bayes_eff}.
In addition, events containing single electrons misclassified as S1 signals (cS1$<$25~PE) are removed. 
The remaining events outside of the ER ROI were removed by the additional selection criteria mentioned above.

\begin{figure}
    \vspace{1mm}
    \centering
    \includegraphics[width=\linewidth]{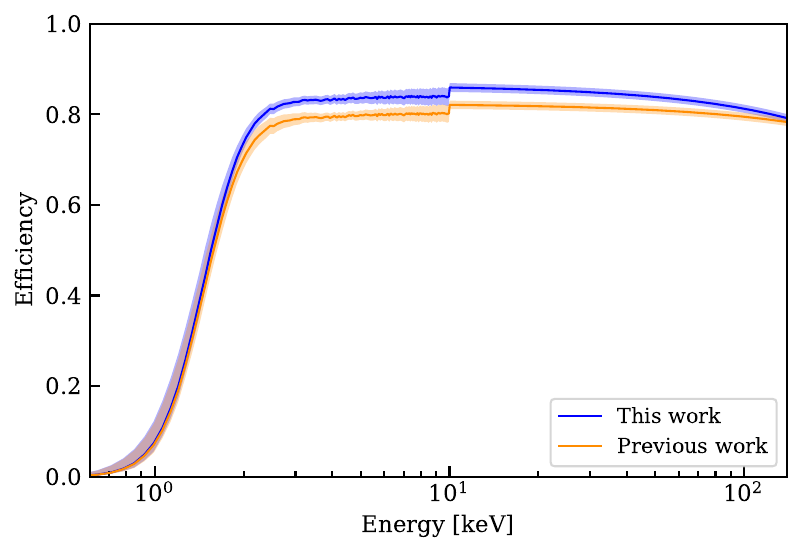}
    \caption{Comparison of total event-selection efficiency between this work and the previous work reported from XENONnT SR0~\cite{Aprile:2022vux}. 
    The step increase at 10~keV in efficiency is due to the nuclear-recoil region of data being blinded below 10~keV during both analyses.
    On average, this method is 3\% more efficient, with the greatest relative improvement in efficiency being in the 2-\mbox{20-keV} energy region.
    }
    \label{fig:total_efficiency}
\end{figure}

The ER event acceptance was calculated using both $^{37}$Ar monoenergetic data within the fiducial volume, and clean $^{212}$Pb data within the fiducial volume from $^{220}$Rn calibration.
The same procedure as in~\cite{Aprile:2022vux} was used for the acceptance calculation, but using solely the S1 and S2 NBC event selection criteria for detector signal quality selection. 
The results of these efficiency calculations are shown in Fig.~\ref{fig:total_efficiency}.
The results of this analysis are shown in combined energy scale, which is a linear combination of cS1 and cS2 values from an event, the calculation of which is described in~\cite{Aprile:2022vux}. 
For ER events, the combined energy is a reconstructed value of the energy deposited by the interaction in the detector.
The overall efficiency is the product of the individual efficiencies from the S1 signal detection (dominant at low energies), the S2 signal detection, the S1 and S2 NBC selection criteria, the accidental coincidence (AC) selection criteria applied in~\cite{Aprile:2022vux}, and the pairing/single-scatter selection criteria applied in~\cite{Aprile:2022vux}. 

The increase in total efficiency relative to the previous work is 3\%. 
The improvement arises from fewer ER events being removed from the data than in the previous method, which relied on several more sequential selection criteria.

The effectiveness of this method at removing events outside of the ER ROI was calculated using the SR0 ER dataset. 
Events outside of the ER ROI were tagged using the set of signal-quality selection criteria developed in~\cite{Aprile:2022vux}. 
These tagged populations were then used to evaluate the relative effectiveness of the S1 and S2 NBC selections at removing events outside of the ER ROI.
The population overlap between events rejected by the previous sequential \mbox{quality-selection} criteria and the S1 and S2 NBC selection criteria of this analysis was 96\%. 
This can be interpreted to mean that the NBC selection criteria are successful at removing events originating from gaslike S1 and S2 signals, misclassified S1 or S2 signals, and misreconstructed signals due to misclustering of individual PMT signals such as dark counts, lone hits, or multiple signals from multiple interactions merged together, populations similarly observed in the $^{220}$Rn calibration dataset. 
The nonunitary overlap between populations removed by the selection methods can be attributed to the increase in ER \mbox{event-selection} efficiency. 

\subsection{Analysis of electronic-recoil spectrum}

\begin{figure}
    \vspace{3mm}
    \centering
    \includegraphics[width=\linewidth]{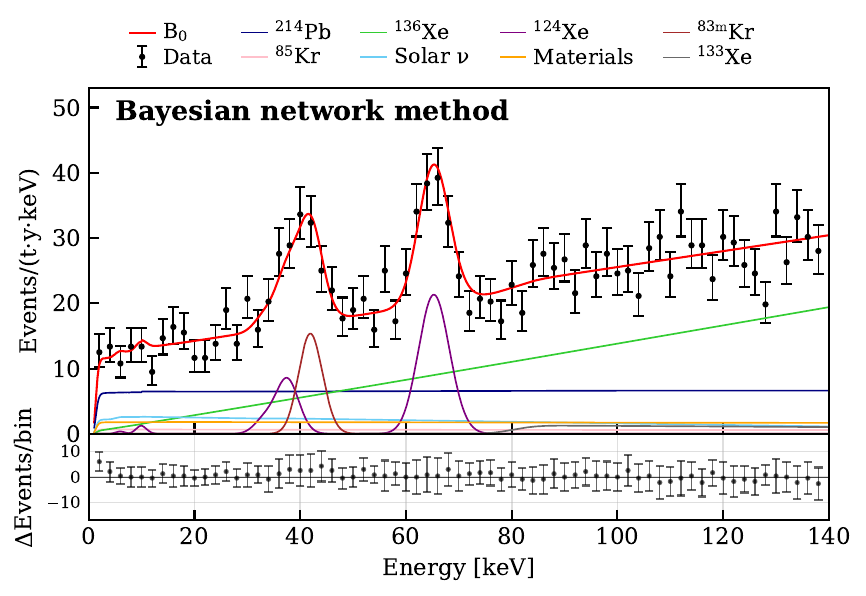}
    \caption{Top panel: SR0 ER data (black) selected using the Bayesian network-based approach with the best-fit background-only model $B_0$ (red). 
    The subdominant AC contribution is not visible. 
    Bottom panel: $\Delta$events/bin corresponds to the difference in observed data in events/(tonne$\cdot$year$\cdot$keV) between those selected using the Bayesian network-based approach and those selected using the previous method described in~\cite{Aprile:2022vux}. 
    The increase of observed events particularly at energies below 20~keV is consistent with the increase in efficiency of this work relative to~\cite{Aprile:2022vux}.} 
    \label{fig:fit_0_140_kev_nbc}
\end{figure}

The low-energy electronic-recoil spectrum of XENONnT SR0 was measured in a total exposure of 1.16~tonne-years; it mainly consists of radiogenic ER events from \textsuperscript{222}Rn contamination in the LXe target. 
The spectrum is shown in Fig.~\ref{fig:fit_0_140_kev_nbc} and was fit by a detailed background model, with all components as described in~\cite{Aprile:2022vux}, with a step-function approximation to account for the electron binding energies in the solar neutrino spectrum as suggested in \cite{amaral2023direct}, and by using an unbinned \mbox{maximum-likelihood} framework.
The goodness-of-fit measurement for the spectrum is $\chi^2/N_{\text{dof}}\ = 128.64/128 = 1.004$ (\textit{p} value of 0.467).
As in~\cite{Aprile:2022vux}, the ER data are consistent with the \mbox{background-only} hypothesis. 

The individual background components contributing to the SR0 ER dataset are summarized in Table~\ref{tab:rate_components}. 
It should be noted that the uncertainties on each component in Table~\ref{tab:rate_components}, while valid for each study individually, are strongly correlated between this work and the previous work.
The dominant source of ER events at low energies is the $\beta$ decay of $^{214}$Pb. 
The activity concentration of $^{214}$Pb in SR0 from this analysis was estimated to be $(1.39\pm0.08)$~µBq/kg, which agrees with the best-fit activity concentration of $^{214}$Pb reported in~\cite{Aprile:2022vux}.
The higher number of observed events in this work can be attributed to the higher average selection efficiency. 
The lower-fitted contribution of $^{133}$Xe to the total number of events, which is a second-order contribution to the event rate above 80~keV, can be attributed to the decrease in relative efficiency gain above 80~keV seen in Fig.~\ref{fig:total_efficiency}. 
The remaining components' best-fit number of observed events are within the expected uncertainty ranges of the reported values in~\cite{Aprile:2022vux}.
 
\begin{table}
    \centering
    \begin{tabular}{|c||c|c|}
    \hline 
    Component & Fit (this work) & Fit (prev. work) \\ \hline
         $^{214}$Pb & 1050 $\pm$ 130 & 960 $\pm$ 120 \\ 
         \hline
         $^{85}$Kr & 100 $\pm$ 60 & 90 $\pm$ 60 \\ 
         \hline
         Materials & 280 $\pm$ 50 & 270 $\pm$ 50 \\ 
         \hline
         $^{136}$Xe & 1580 $\pm$ 60 & 1550 $\pm$ 50 \\ 
         \hline
         Solar $\nu$ & 310 $\pm$ 30 & 300 $\pm$ 30 \\ 
         \hline
         $^{124}$Xe & 250 $\pm$ 30 & 250 $\pm$ 30 \\ 
         \hline
         AC & 0.71 $\pm$ 0.03 & 0.71 $\pm$ 0.03 \\
         \hline
        $^{133}$Xe & 80 $\pm$ 60 & 150 $\pm$ 60 \\
         \hline
          $^{83\text{m}}$Kr & 101 $\pm$ 17 & 80 $\pm$ 16 \\ 
         \hline    
    \end{tabular}
    \caption{Best-fit background model, $B_0$ components, with number of events observed for each component in SR0 within the energy range of (1, 140) keV. The right column has the number of events reported from~\cite{Aprile:2022vux}.}
    \label{tab:rate_components}
\end{table}

Additional potential sources of systematic uncertainty in the results of this analysis include any introduced by the NBC method and variations in signal shapes between training and experimental data.
We found that the choice of number of attributes used in the NBC did not cause a statistically significant difference in the deterministic classification results shown in Table~\ref{table:classification}.
For the signal size, the most significant systematic effects in this work would arise from nonuniformities in the drift field within the fiducial volume, which were already accounted for in the reconstructed combined energy resolution. 
Thus, to avoid double counting, the systematic uncertainties are calculated using a method identical to that in \cite{Aprile:2022vux}.

\section{Conclusion and future work}\label{sec:conclusions}

We demonstrated the first use of a Bayesian network to perform detector signal classification, waveform-based event selection, and subsequent analysis in a full-scale dual-phase Xe TPC. 
Applying the NBC-based metric in event selection reduces significantly detector signals outside of the region of interest and solves the need for additional selection criteria based on detector signal quality. 
The method can be used both as an independent analysis framework and as a valuable cross-check for analyses that use lower-dimensional features in subsequent event selection steps. 
The development of this method relies on simulated data, and the selection criteria rely on calibration data, without optimizing on ER search data. 
Therefore, due to the NBC being developed blinded to the ER dataset, it is agnostic to any new interactions present in the XENONnT SR0 ER search data, allowing us to demonstrate that the Bayesian network-based method of signal selection corroborates the ER background-only hypothesis of the low-energy ER data from XENONnT SR0~\cite{Aprile:2022vux}.

In the future, developing data selection criteria primarily using Bayesian network-based methods could help to increase signal-to-background ratios in dark matter detector experiments, thereby increasing experimental sensitivity to new physical processes.
Relaxing the threefold PMT coincidence requirement for a valid S1 signal, which is the primary reason for efficiency loss below 2~keV, could be viable without a significant increase in the accidental coincidence rate by using a Bayesian network specially trained for classifying these events.
With this additional improvement on efficiency in a future study dedicated to reducing accidental coincidence background rates, low-energy physics phenomena could be probed to new sensitivities. 

Other future directions of this work in XENONnT's physics program include the use of dynamic Bayesian networks~\cite{DeanKanazawa1989}, which incorporate temporal dependencies between the attribute nodes. 
If the use of dynamic Bayesian networks adequately accounts for the temporal dependencies, then the output of the probability queries of the model will be suitable for use in an end-to-end probabilistic analysis.
One such meritable analysis would be to define the fiducial volume based not upon event localization algorithms, but on the posterior probabilities from a Bayesian network.

\section*{Acknowledgments}
We gratefully acknowledge support from the National Science Foundation, Swiss National Science Foundation, German Ministry for Education and Research, Max Planck Gesellschaft, Deutsche Forschungsgemeinschaft, Helmholtz Association, Dutch Research Council (NWO), Weizmann Institute of Science, Israeli Science Foundation, Binational Science Foundation, Fundacao para a Ciencia e a Tecnologia, R\'egion des Pays de la Loire, Knut and Alice Wallenberg Foundation, Kavli Foundation, JSPS Kakenhi and JST FOREST Program in Japan, Tsinghua University Initiative Scientific Research Program and Istituto Nazionale di Fisica Nucleare. 
This project has received funding/support from the European Union's Horizon 2020 research and innovation programme under the Marie Sk\l{}odowska-Curie Grant Agreement No 860881-HIDDeN. 
Data processing was performed using infrastructures from the Open Science Grid, the European Grid Initiative and the Dutch national e-infrastructure with the support of SURF Cooperative. 
We are grateful to Laboratori Nazionali del Gran Sasso for hosting and supporting the XENON project.

\clearpage

\onecolumngrid
\section*{Appendix}\label{appendix}

\begin{figure*}[h!]
  \includegraphics[width=0.32\textwidth]{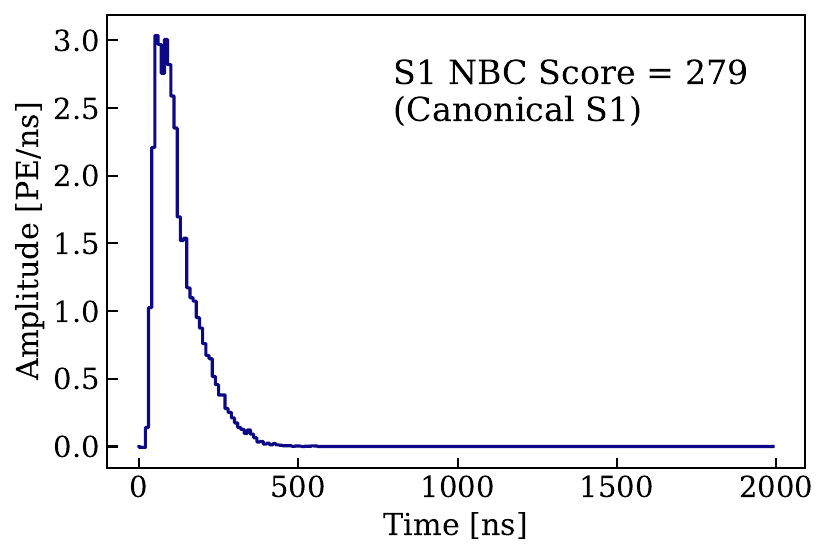}
  \includegraphics[width=0.32\textwidth]{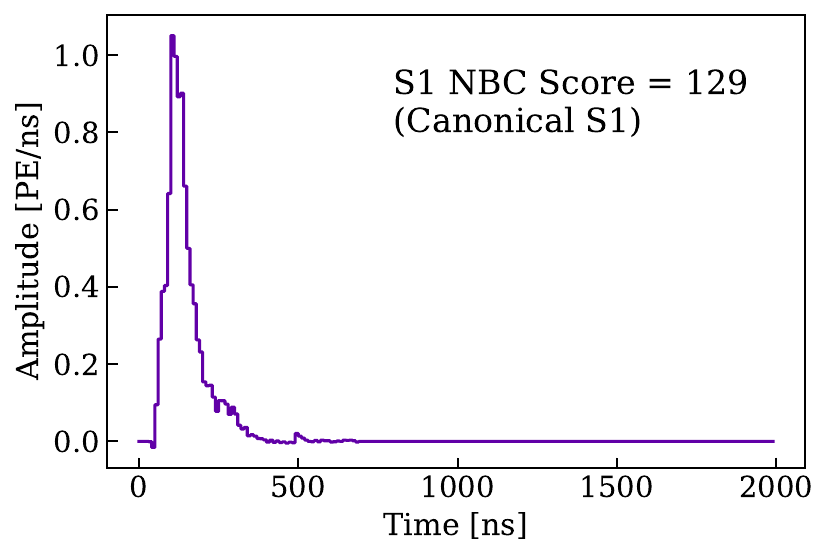}
  \includegraphics[width=0.32\textwidth]{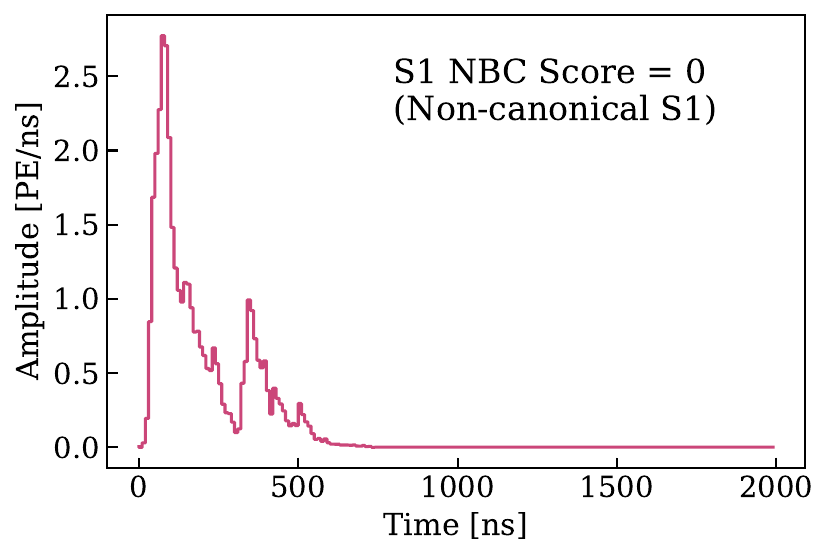}

  \caption{Observed S1 signals from the \mbox{32.1-keV} (left), \mbox{9.4-keV} (middle), and merged \mbox{41.5-keV} (right) decays of \textsuperscript{83m}Kr during SR0 calibration. 
  In particular, the \mbox{41.5-keV} peak has an S1 NBC score of 0, indicating that it is neither a canonical S1 nor a canonical S2. 
  In reality, this is a merged double S1. 
  The NBC scores between the three populations allow for selection of the merged waveforms based solely on shape. 
  The NBC selection criteria described in Sec.~\ref{sec:signal_characterization} would ensure that such double S1s were rejected in ER event selection for analysis. 
  The color of each waveform corresponds to the NBC score as shown in Fig.~\ref{fig:kr83m_nbc}.
}
\end{figure*}

\begin{figure*}[h]      
\includegraphics[width=0.32\textwidth]{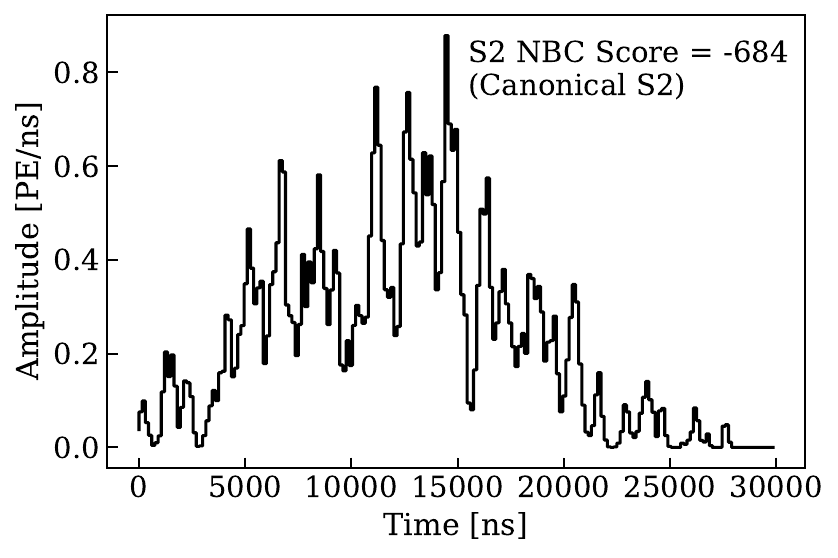}
\includegraphics[width=0.32\textwidth]{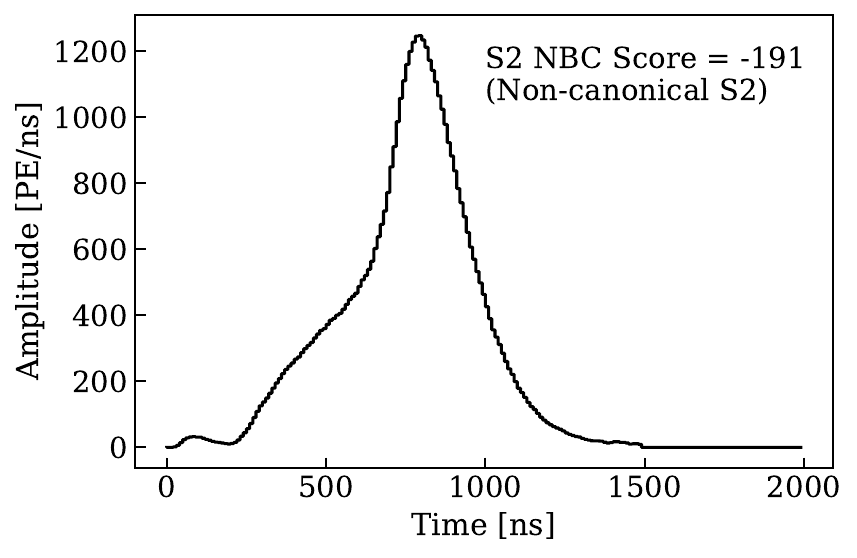}
\includegraphics[width=0.32\textwidth]{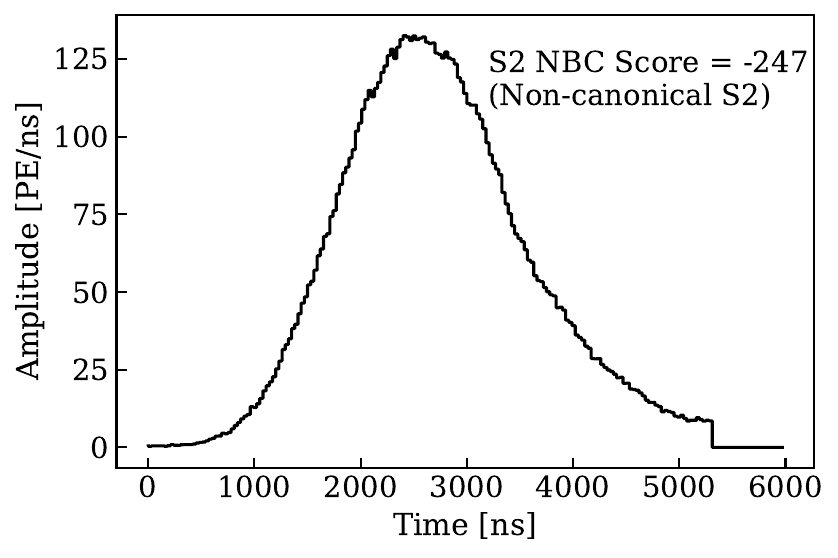}
  \caption{Observed S2 signals from \textsuperscript{83m}Kr calibration. 
  A canonical S2 signal (left) produced from \textsuperscript{83m}Kr decay passes the NBC selection criteria. 
  A high-energy interaction in the GXe region of the detector produced a merged, noncanonical S2-classified signal (middle), and this signal is vetoed by the NBC selection criteria. 
  Finally, an ionization signal produced in gas (right) is also rejected by the NBC selection criteria. 
  Robust rejection of such events is important both for proper ER event selection, and for calibration and efficiency calculations.  
}
\end{figure*}

\twocolumngrid 
\bibliography{references.bib}

\end{document}